# Effectiveness of nanoinclusions for reducing bipolar effects in thermoelectric materials

Samuel Foster[*] and Neophytos Neophytou

School of Engineering, University of Warwick, Coventry, CV4 7AL, UK

[*]S.Foster@warwick.ac.uk

## Abstract

Bipolar carrier transport is often a limiting factor in the thermoelectric efficiency of narrow bandgap materials (such as $Bi_2Te_3$ and PbTe) at high temperatures due to the introduction of an additional term to the thermal conductivity and a reduction in the Seebeck coefficient. In this work, we present a theoretical investigation into the ability of nanoinclusions to reduce the detrimental effect of bipolar transport. Using the quantum mechanical non-equilibrium Green's function (NEGF) transport formalism, we simulate electronic transport through two-dimensional systems containing densely packed nanoinclusions, separated by distances similar to the electron mean-free-path. Specifically, considering an n-type material, where the bipolar effect comes from the valence band, we insert nanoinclusions that impose potential barriers only for the minority holes. We then extract the material's electrical conductivity, Seebeck coefficient, and electronic thermal conductivity including its bipolar contribution. We show that nanoinclusions can indeed have some success in reducing the minority carrier transport and the bipolar effect on both the electronic thermal conductivity and the Seebeck coefficient. The benefits from reducing the bipolar conductivity are larger the more conductive the minority band is to begin with (larger hole mean-free-path in particular), as expected. Interestingly, however, the benefits on the Seebeck coefficient and the power factor are even more pronounced not only when the minority mean-free-path is large, but when it is larger compared to the majority conduction band mean-free-path. Finally, we extract an overall estimate for the benefits that nanoinclusions can have on the *ZT* figure of merit.

**Keywords:** Thermoelectrics, nanotechnology, nanoinclusions, modeling and simulation, quantum transport, NEGF, Seebeck coefficient, bipolar effects.

# I. Introduction

Thermoelectric (TE) materials, which convert between heat and electrical energy, have drawn increased attention in recent years due to the development of a plethora of advanced materials such as half-heuslers, skutterudites, oxides, clathrates, etc., and especially their nanostructured forms with record-low thermal conductivities,[1,2,3,4,5,6,7,8,9] accompanied by the emergence of high-throughput material search calculations and databases.[10,11] The performance of TE materials is quantified by the figure of merit $ZT = \sigma S^2 T/(\kappa_l + \kappa_{el,tot})$ where $\sigma$ is the electrical conductivity, $S$ is the Seebeck coefficient, $T$ is the temperature, $\kappa_l$ is the lattice thermal conductivity, and $\kappa_{el,tot}$ is the total electronic thermal conductivity composed of $\kappa_{el,e}$, the contribution due to electrons, $\kappa_{el,h}$, the contribution due to holes, and $\kappa_{bi}$, the bipolar thermal conductivity. These new generation TE materials allowed a large increase in $ZT$ from 1 to >2.5 in recent years due to drastically lower lattice thermal conductivities.[9]

Many traditional and contemporary TE materials, however, suffer from narrow bandgaps (e.g. PbTe ~ 0.3 eV,[12] $Bi_2Te_3$ ~ 0.2 eV,[13] SnSe ~ 0.39-0.8 eV),[14,15] making them susceptible to bipolar effects at high operating temperatures. In this case: i) the overall electronic thermal conductivity $\kappa_{el,tot}$ increases due to contributions from both electrons and holes, ii) an additional term, the bipolar thermal conductivity $\kappa_{bi}$ is introduced (a result of electron-hole recombinations at the contacts),[16] which also causes large increases in the Lorenz number,[17] and iii) the overall Seebeck coefficient drops as both electrons and holes contribute to it with opposite signs. In order to reduce the bipolar effect, minority carriers need to be blocked, e.g. using heterostructure designs,[18,19] band engineering to widen the bandgap,[20,21] grain boundaries with barriers for minority carriers,[22] etc.

One of the most commonly employed methods for the reduction in the lattice thermal conductivity, $\kappa_l$, on the other hand, has been the use of nanoinclusions (NIs),[23,24,25,26,27,28] which could be experimentally easier to realize compared to heterostructures. NIs are highly effective at reducing the lattice thermal conductivity $\kappa_l$ arising due to phonon transport. At high temperatures, however, $\kappa_l$ is typically reduced naturally due to enhanced phonon-phonon scattering. In the most widely used thermoelectric materials it is the emergence of the $\kappa_{bi}$, a term that arises from electronic transport, which adds a significant contribution to the total thermal conductivity and degrades performance. Since NIs also influence the electronic transport, it seems pertinent to ask whether they can also reduce the bipolar effect, and whether in that way it is possible to achieve simultaneous reductions in $\kappa_l$, $\kappa_{el,tot}$, and $\kappa_{bi}$. To accurately examine this, however, advanced simulations are needed, which capture all geometrical complexities (non-uniform structures, mixing low-dimensional NIs with higher-dimensional bulk regions), as well as all transport physics at the nanoscale (quantum reflections, interferences, quantization, quantum tunnelling etc.).[29]

In this work we use the quantum mechanical non-equilibrium Green's function (NEGF) formalism, including electron-phonon interactions, which can accurately capture the important details specified above, to calculate the thermoelectric transport coefficients of bipolar systems embedded with nanoinclusions. We identify the conditions for which the bipolar effect is most significant, and those for which nanoinclusions are most effective at reducing it. We then employ literature thermal conductivity data for a series of materials to estimate the overall improvement in the *ZT*. The paper is organised as follows: In Section II we explain the simulation method we employ and describe the geometries we consider. In Section III we present and discuss our results, and finally in IV we conclude.

## II. Approach

We employ the NEGF formalism to compute electronic transport, including interactions of electrons with acoustic phonons (acoustic phonon scattering). The theory and computational details can be found in various places in the literature,[30,31,32] so are not included here. The simulated system is a 2D channel described using the effective mass approximation. The effective mass is considered uniform throughout the channel, including for the nanoinclusions, although this mass is altered to consider non-symmetric conduction and valence bands. The nanoinclusions are modelled as cylindrical potential energy barriers in the valence band as shown in Fig. 1. We consider barrier heights of $V_B$

= 0.2 eV and barrier diameters of $d$ = 3 nm. The system dimensions we simulate are width $W$ = 30 nm and length $L$ = 60 nm. Acoustic phonon scattering is considered and the strength of the electron-phonon coupling is such that an electronic mean-free-path (MFP) of 15 nm at $T$ = 300 K is achieved, meaning we have diffusive transport in the channel.[33,34] Note that from scattering theory, we know that the rate is proportional to the temperature as: $1/\tau = \pi D_A^2 k_B T g(E)/\hbar c_l$ where $D_A$ is the acoustic deformation potential, $k_B$ is the Boltzmann constant, $g(E)$ is the density of states, $\hbar$ is the reduced Planck constant, and $c_l$ is the material's elastic constant.[35] Thus, our calibrated MFP of 15 nm only holds at $T$ = 300 K and will decrease linearly with increasing temperature in the simulations we consider below. In all cases, however, the distance between NIs, as shown in Fig. 1, is similar to our nominal electron-phonon scattering MFP at room temperature. We select this on purpose so that NIs have a significant influence on the transport properties.

For each case considered, the conduction and valence bands are simulated separately (i.e. we simulate a single band in NEGF) and then combined as described below. The total electrical conductivity, total Seebeck coefficient, total electronic thermal conductivity, individual band electronic thermal conductivities, and bipolar thermal conductivity, are evaluated by:[17,19]

$$\sigma = \sigma_e + \sigma_h \tag{1}$$

$$S = \frac{\sigma_e S_e + \sigma_h S_h}{\sigma_e + \sigma_h} \tag{2}$$

$$\kappa_{el,tot} = \kappa_{el,e} + \kappa_{el,h} + \kappa_{bi} \tag{3}$$

$$\kappa_{el,e/h} = \frac{1}{q^2 T}\int \sigma(E)(E - E_F)^2 dE - \sigma_{e/h} S_{e/h}^2 T \tag{4}$$

$$\kappa_{bi} = \frac{\sigma_e \sigma_h}{\sigma_e + \sigma_h}(S_e - S_h)^2 T \tag{5}$$

where the subscripts 'e' and 'h' indicate the contributions from electrons and holes respectively, $q$ is the elementary charge, $E$ is energy, and $E_F$ is the Fermi level. Note that $S_e$ and $S_h$ have opposite signs and therefore the total Seebeck coefficient, $S$, is reduced when both bands contribute to the transport, normally also reducing the power factor despite any gains in $\sigma$. Similarly, $\kappa_{bi}$ is maximized when both bands contribute to carrier transport, often when the Fermi level is placed in the midgap. It should also be noted that $\kappa_{bi}$ is a conductivity-limited quantity, and is primarily determined by the conductivity of the minority carrier (the smaller of the two as seen by the right term in Eq. (5)).[36] This can be seen from its physical origin of electron-hole recombinations in the contacts.[16] Increasing an already large flow of electrons will not increase electron-hole recombinations if there are no further holes to recombine with, but an increase in hole flow would produce a similar increase in the electron-hole recombination rate. Note, also, that although the conductivities $\sigma$ and $\kappa$ are used in the equations above, due to the 2D nature and finite channel length of our NEGF simulations, we extract the conductances, $G$ and $K$, and it is understood in the analysis that follows that $\sigma$ and $\kappa$, and $G$ and $K$ can be used interchangeably.

We consider a matrix material with bandgap $E_g = 0.2$ eV, similar to that of $Bi_2Te_3$ for example. We place the Fermi level in alignment with the conduction band edge as this will provide optimal power factors, at least in a unipolar material.[33] We then insert a dense network of nanoinclusions as in Fig. 1 by the introduction of potential barriers in the valence band, but allow perfect band alignment in the conduction band (a successful power factor and $ZT$ improvement strategy outlined in Ref. 37 for the case of SrTe inclusions in PbTe). Just by looking at Eqs. (1)-(5), it is obvious that the higher the hole conductivity, then: i) the higher the overall electronic conductivity, ii) the lower the overall Seebeck coefficient, and iii) the higher the bipolar conductivity. We therefore would like to answer the question: how effective are nanoinclusion induced barriers in the valence band at reducing bipolar effects? In addition, in the light of the complexity in the bandstructures of new generation TE materials: under what bandstructure conditions is the use of such nanostructuring most effective?

## III. Results and Discussion

To see the effect of bandstructure in a simplified manner, we begin our investigation by considering the influence of different effective masses, lighter and heavier (which provide different electronic mean-free-paths and conductivities) and the influence of valence band nanostructuring on the bipolar and electronic thermal conductivity. Figures 2(a)-(b) show the four bandstructures we consider with effective masses as follows: $m_c = m_0$, $m_v = m_0$ (green-cross lines); $m_c = m_0$, $m_v = 0.5m_0$ (red-square lines); $m_c = 0.5m_0$, $m_v = m_0$ (blue-star lines); $m_c = 0.5m_0$, $m_v = 0.5m_0$ (black-circle lines), where $m_c$ is the conduction band effective mass, $m_v$ is the valence band effective mass, and $m_0$ is the electron rest mass. In each case the Fermi level coincides with the conduction band minimum, i.e. $E_F = 0$ eV (red-dashed line in Fig. 2(a)). Our intent here is to explore the qualitative effect of conduction/valence band symmetry and asymmetry in bipolar transport, i.e. with respect to different combinations of light/heavy effective masses.

In Fig. 2(c)-(d), we show the bipolar electronic thermal conductivity, $\kappa_{bi}$, versus temperature for each bandstructure. In each case we consider structures without NIs (solid lines) and with NIs (dashed lines), which introduce barriers in the valence band. For all bandstructure examples $\kappa_{bi}$ increases with temperature, as expected, due to the broadening of the Fermi distribution that begins to pick up carriers from the valence band. The largest bipolar effect can be seen in the high conduction cases when the effective masses are smaller. Particularly important is the influence of a light valence band (red and black solid lines). This is because (as shown earlier by Eq. (5)) the conduction of the minority band, which has a significantly smaller value, is dominant in determining $\kappa_{bi}$. Upon nanostructuring (dashed lines), the relative reduction in $\kappa_{bi}$ however, is also larger in the case where the masses are smaller and will be discussed in more detail later.

In Fig. 2(e)-(f) we show the corresponding total electronic thermal conductivities, $\kappa_{el,tot}$. Again all bandstructures show an increase in $\kappa_{el,tot}$ with temperature. At low temperatures $\kappa_{el,tot}$ is dominated by $\kappa_{el,e}$, the contribution from the conduction band, since we are close to degenerate carrier concentration conditions ($E_F$ is at $E_C$). At higher temperatures, however, the contribution of $\kappa_{bi}$ becomes important, fuelled mostly from light valence bands, which add more significantly to minority carrier transport. Thus, in the light band case of Fig. 2(e) (black line) a larger increase in $\kappa_{el,tot}$ is observed. Likewise, in the cases of asymmetric masses between the conduction and valence band in Fig. 2(f), although the *heavy*-conduction-*light*-valence bandstructure (red-solid line) starts lower compared to the *light*-conduction-*heavy*-valence bandstructure case (blue-solid line), at higher temperatures the two merge. With the introduction of NIs, both the $\kappa_{bi}$ and $\kappa_{el,tot}$ are reduced (dashed lines in Fig. 2), but the light valence mass cases are reduced much more, something we discuss below in more detail.

We now proceed in similarly analyzing the effect of the different band masses and of valence band nanostructuring on the rest of the thermoelectric coefficients: the conductance, $G$, the Seebeck coefficient, $S$, and the power factor, $PF$ defined as $GS^2$, versus temperature (shown in Fig. 3). We use the same four bandstructures shown in Fig. 2(a)-(b). Since we are close to degenerate carrier concentration conditions, $G$ is dominated in all cases by the effective mass of the conduction band (meaning the NIs on the valence band have little influence on $G$) and shows a small increase with temperature (Fig. 3(a)-(b)). Note that we use a fixed Fermi level ($E_F = 0$ eV), so the increase in phonon scattering that would normally reduce conductance with increasing temperature is offset by an increase in the carrier concentration due to the broadening of the Fermi distribution. The Seebeck coefficient, however, slowly decreases with temperature since the valence band (which has opposite Seebeck sign) exponentially increases its contribution to transport (Fig. 3(c)-(d)). The reason the influence of temperature is more significant on $S$, is that although $\sigma_e$ dominates $\sigma_h$, $\sigma_e S_e$ is relatively less dominant over $\sigma_h S_h$ as seen in Fig. 4, which shows that the ratios of $G_e/G_h$ are significantly higher compared to the ratios of $G_e S_e/G_h S_h$. Therefore, from Eq. (1) and Eq. (2) it can be seen that the valence band has more influence on the total Seebeck than on the total conductance. Consequently, the introduction of NIs (dashed lines) on the valence band limits the contribution to transport and $S$ is somewhat recovered (compare the dashed to the solid lines in Fig. 3(c)-(d)). These effects can be seen in the power factors in Fig. 3(e)-(f) as well, where the power factors partially recover with the introduction of NIs due to this recovery of $S$.

We now quantify the changes we observe in $\kappa_{bi}$, $\kappa_{el,tot}$, and the $PF$ due to the introduction of NIs on the valence band for the four bandstructures shown in Fig. 2(a)-(b). Figure 5 shows the percentage changes in these quantities with temperature. In Fig. 5(a), for the $\kappa_{bi}$, it can be seen that at 300 K an initial reduction in $\kappa_{bi}$ close to 80% is observed for the light valence band mass materials (red/black lines), and around 60% for the heavier mass ones. The percentage generally decreases with temperature, indicating that NIs are less effective in reducing $\kappa_{bi}$ as the temperature increases. The reason for this can be seen from the NEGF resolved transmissions of the valence band (alone), the quantities we use to extract the valence band TE coefficients, which we show in Fig. 6. Note that the energy values are relative to the valence band edge (although plotted on the positive x-axis) and increase in value moving into the band. While the transmission of the pristine valence band reduces with increasing temperature due to the increasing phonon scattering rate previously mentioned (solid lines), the transmission of the structure with NIs remains largely unchanged with increasing temperature. This suggests that the scattering mechanism introduced by the NIs (in combination with phonon scattering) is somewhat less dependent on temperature at first order. As the temperature increases, the electron-phonon scattering limited transmission (solid lines) and the transmission of the valence band with NIs plus phonon scattering (dashed lines) begin to converge. This means the introduction of the NIs has a smaller effect on the valence band conductance at higher temperatures. The reduction of the percentage decrease in $\kappa_{bi}$ can also be accounted for by $\kappa_{bi}$ being dominated by the minority carrier conductivity, $\sigma_h$, as explained earlier. At lower temperatures $\sigma_h$ is small compared to $\sigma_e$, and therefore dominates $\kappa_{bi}$ (i.e. the right term in Eq. (5) changes more for smaller $\sigma_h$ values). At higher temperatures $\sigma_h$ begins to approach $\sigma_e$ and so the reduction in $\sigma_h$ from the NIs does not have as large a relative impact on $\kappa_{bi}$. However, this is only a relative effect - the absolute value of the $\kappa_{bi}$ reduction at high $T$ is much larger.

Despite this relative reduction in the influence of the NIs on $\kappa_{bi}$, the overall impact on $\kappa_{el,tot}$ increases with temperature as shown in Fig. 5(b). At lower temperatures $\kappa_{el,tot}$ is dominated by the conduction band $\kappa_{el,e}$ so reducing $\kappa_{bi}$ does not have a significant impact. At larger temperatures, however, $\kappa_{bi}$ becomes more significant (and becomes an increased proportion of $\kappa_{el,tot}$) and reducing it therefore has a larger impact on $\kappa_{el,tot}$. This effect is, however, counteracted by the NIs falling impact on $\kappa_{bi}$ shown in Fig. 5(a), and therefore the decrease begins to saturate at higher temperatures. Likewise, in Fig. 5(c) we see that the *PF* is also increasingly improved by the introduction of NIs with temperature in all bandstructure cases, an effect attributed to the increase in *S*. In a similar case to $\kappa_{el,tot}$, the valence band contribution to $S$ becomes increasingly more significant at higher temperatures, and thus the influence of the NIs in reducing $S_h$ increases likewise. The relative increase in the PF versus temperature between the pristine materials and those with NIs is continuously increasing (reaching even up to ~20% at 600 K), without any signs of saturation up to the temperatures we consider (in contrast to $\kappa_{bi}$ and $\kappa_{el,tot}$). We stress that PF improvement value of 20% are quite significant for TE materials, which usually experience only incremental changes upon the introduction of new concepts. The continuous improvement with temperature is because, as seen in Fig. 4, the quantity $\sigma_e S_e$ in Eq. 2 (for the overall Seebeck coefficient) becomes less dominant over $\sigma_h S_h$ as the temperature increases. This maintains the influence of the valence band (and consequently NIs) at higher temperatures for the overall Seebeck coefficient and the *PF*. Note the multiple effects at play here: $\kappa_{bi}$ is determined primarily by the valence band due to Eq. 5 being multiplicative, but its influence decreases with temperature due to two reinforcing effects: i) $\sigma_h$ begins to approach $\sigma_e$ so the valence band (and the NIs) becomes less influential, and ii) the electron-phonon scattering increases (as acoustic phonon scattering rate is ~$T$), which makes the NIs relatively less influential as a scatterer. The *PF* on the other hand is determined primarily by the conduction band since Eqs. 1 and 2 are additive, and its influence increases with temperature according to two competing effects: i) $\sigma_h$ begins to approach $\sigma_e$ and $\sigma_h S_h$ is begins to approach $\sigma_e S_e$ so the valence band (and the NIs) become more influential, and ii) as before, the electron-phonon scattering increases, which makes the NIs become less influential. For the temperatures that we have considered, the first of these effects is dominant and the *PF* benefits due to the NIs increase with increasing temperature without saturating, but it could be possible that there is an upper limit at higher temperatures.

An important observation here, is that the highest overall *PF* improvement is not found in the case where the band masses (especially that of the valence band) are low, i.e. their conductivity is high. The highest benefits on the *PF* of nanostructuring the valence band to reduce the bipolar effect come in the case where the valence/conduction band effective masses are asymmetric, with the valence mass larger than the conduction mass. In this case, with a factor of 2 asymmetry in the masses, the *PF* improvements at 600 K are nearly double compared to the symmetric low masses case. Despite the fact that the reduction in $\kappa_{bi}$ is dominated by the valence band conductivity (black/red lines and green/blue lines of same valence band conductivity overlap in Fig. 5(a)), in the PF case, the more comparable to $\sigma_e S_e$ the term $\sigma_h S_h$ is, the largest the relative benefits as $\sigma_h S_h$ is decreased as seen in Eq. 2. A conduction band with lower conductivity, will have a lower $\sigma_e S_e$, which will be closer in value to $\sigma_h S_h$, which is typically low anyway. This leads to the non-intuitive observation that it is not only materials with a light mass and long MFP minority band that are benefitted by nanostructuring, but materials with highly asymmetric heavy-conduction and light-valence bands can experience even higher relative *PF* improvements when bipolar conduction is degraded. The larger benefit from this asymmetry holds for the electronic thermal conductivity components (compare red versus black lines in Fig. 5(b)), but even more importantly for the *PF* (compare red versus black lines in Fig. 5(c)). This also leads to the encouraging result that bipolar materials with rather lower power factors to begin with due to lower majority band conductivity, could have decent chances in recovering from bipolar effect degradation, provided the minority carriers have longer mean-free-paths.

An interesting point we wish to elucidate here a bit further, is the reason that the lighter valence band masses produce the largest reductions in $\kappa_{bi}$ and $\kappa_{el,tot}$ as well as the largest increases in *PF* in the presence of NIs. This can be intuitively understood from simple scattering considerations, where the total MFP for minority carriers in the valence band with NIs can be obtained using Matthiessen's rule:

$$\frac{1}{\lambda_{tot}} = \frac{1}{\lambda_{ph}} + \frac{1}{\lambda_{NI}} \tag{6}$$

where $\lambda_{tot}$ is the MFP of the system, $\lambda_{ph}$ is the MFP due to phonon scattering only, $\lambda_{NI}$ is the effective distance between NIs that the carriers travel before they scatter, and we have assumed that the electron velocity is the same between all cases. To extract an analytical estimate of the MFP due to NIs alone, we determine the number of collision (scattering) events, $N_{coll}$, per unit length (along the transport direction). The inverse of the number of interface scattering events per unit length provides an effective scattering distance $\lambda_{NI}$ between the pores ($\lambda_{NI} = 1/N_{coll}$).[38] Using $\lambda_{NI} = 2d/3\varphi$, as defined in Ref. 38 where $\varphi$ is NI density (~12 % in our system), we find $\lambda_{NI}$ = 16 nm, which is similar to the spacing of our NIs in the simulator as expected. This MFP is comparable to the phonon scattering MFP in the nominal, heavier band case we consider with $m_v = m_0$ (~ 15 nm) that we have calibrated to in our simulator. In the light band case, however, with $m_v = 0.5m_0$, $\lambda_{ph}$ is doubled since the carrier velocities are higher ($v \propto m^{-1/2}$) and the phonon scattering times are longer ($\tau \propto m^{-D/2}$ where $D$ is dimensionality, in our case $D=1$ as our structures are narrow enough to be composed of individual 1D subbands).[35] Thus, $\lambda_{light\_mass} \propto v\tau \propto m^{-1/2} m^{-D/2} = m^{-1} \propto 2\lambda_{heavy\_mass}$. Therefore, from Eq. (6), the NIs are more influential in the light band case in achieving larger relative reduction in minority carrier transport, i.e., $\lambda_{tot} = 7.7$ nm in the heavier band case (halved), while it falls by two thirds from $\lambda_{ph} = 30$ nm to $\lambda_{tot} = 10$ nm in the lighter band case. Thus, in order to achieve maximum benefits, NIs should be placed on the order of the MFP of the material's minority carriers, or even more densely.

Such expectations are evident in our NEGF simulations in the transmission functions, and further details are captured as well. The transmission of the valence bands with masses: $m_v = m_0$ (green lines) and $m_v = 0.5m_0$ (black lines) for the cases without (solid lines) and with (dashed-dotted lines) NIs shown in Fig. 7. Again, the energy values are relative to the valence band edge (although plotted on the positive x-axis) and increase in value moving into the band. It can be seen that at lower electron energies the lighter mass (black lines) is more affected by the introduction of NIs leading to a greater reduction in the hole conductance – in fact it seems that the NIs dominate the transmission of both heavy and light bands, somewhat more than expected. For example, at lower energies we see a deviation from Matthiessen's rule with the transmission in the presence of NIs being the same for short and longer electron MFPs. This is possibly due to coherent/wave effects at such nanoscale features that are captured within the use of NEGF, and/or because NEGF captures the geometry of the channel more accurately than simple analytical models. Note that the doubling of $\lambda_{ph}$ in the light band case is also captured by our NEGF simulations. This can be observed by plotting the number of modes versus energy for the two valences bands of masses $m_v = m_0$ (green line) and $m_v = 0.5m_0$ (black line) in Fig. 7(b). Using the relation $T(E) = M(E)\langle\lambda(E)\rangle/L$ it can be seen that the ratio of the MFPs of the two systems is equal to their individual $T(E)/M(E)$ ratio.[39] This is just the ratio of the dashed-dotted lines in Fig. 7(a) to the ones in Fig. 7(b), which is shown in the inset of Fig. 7(b), indicating that the light mass material has indeed double the MFP (i.e. ~ 30 nm). Combining our results from Figs. 6 and 7 we can see that the NIs appear to dominate the transmission for both masses and at all temperatures, resulting in very similar transmissions in all cases.

We now consider the effect that introducing NIs has on the overall figure of merit $ZT$. For this, we combine the electronic transport results given above with lattice thermal conductivity results taken from the literature. Reductions in $\kappa_l$ can vary considerably depending on a wide variety of factors, including the material type, NI material, temperature, and the density of NIs. As it is very difficult to perform a comparison upon all these parameters, here we extract an average value across all parameters using experimental results found in the literature as shown in Fig. 8. That average is an overall 58 % reduction in $\kappa_l$, across materials, temperatures, and densities, and we use this to provide an indication of $ZT$ improvements. Due to the 2D nature of our simulations, we calculate electrical conductance and electronic thermal conductance which are dimensionally incompatible with the lattice thermal conductivity taken from the literature. An initial $ZT$ value is therefore calculated by utilizing experimental results for $\sigma$, $S$, $\kappa_e$, and $\kappa_l$ from the literature (specifically, since our bandgap is most similar to $Bi_2Te_3$ we take a representative set of values taken from Ref. 40 at 500 K) and then we consider the relative change that our simulations will impose on literature data of pristine structures. Therefore, the $ZT$ improvement is then computed as:

$$ZT = \frac{\left(\dfrac{G_{NI}}{G_{pristine}}\right)\sigma_{lit}\left(\dfrac{S_{NI}}{S_{pristine}}\right)^2 S_{lit}^2}{\left(\dfrac{K_{e-NI}}{K_{e-pristine}}\right)\kappa_{e-lit} + \left(\dfrac{\kappa_{l-NI}}{\kappa_{l-pristine}}\right)\kappa_{l-lit}} \quad (7)$$

where $\sigma_{lit}$ etc. are taken from Ref. 40, and the ratios $\dfrac{G_{NI}}{G_{pristine}}$ etc. are taken from this work in the case of $G$, $S$ and $K_e$, and from Fig. 8 for $\kappa_l$. For the $G$, $S$ and $K_e$ ratios we take the results of our simulations for the bandstructure $m_c = 0.5m_0$, $m_v = 0.5m_0$ (black lines in Fig. 5) and for the $\kappa_l$ ratio we take a constant 0.42 (i.e. a 58% reduction) across all temperatures. Note that the bandstructure of $Bi_2Te_3$ is highly anisotropic, thus, what we provide is just a rough indication on how ZT in the case of $Bi_2Te_3$ will benefit in the presence of nanoinclusions.[48]

In Fig. 9 we show the percentage increase in ZT due to the introduction of nanoinclusions, for the nominal material we consider with bandstructure $m_c = 0.5m_0$, $m_v = 0.5m_0$ assuming a constant 58% decrease in $\kappa_l$ (black line). At 300 K the majority of the ZT increase comes from the reduction in $\kappa_l$, however as T increases the improvement increases to almost double due to the NIs impact on the bipolar effect and the electronic properties of the material. With the red star we denote the improvement in ZT calculated using the 80% reduction in $\kappa_l$ at 500 K reported in Ref. 40. Since we take our initial values from here this represents the closest possible estimation to a potential improvement in ZT in a real material (namely $Bi_2Te_3$). This gives us a highly significant 148% increase in ZT. It is expected that this could become even higher as the temperature increases beyond our simulated temperatures due to the NIs increasing impact on the PF and $\kappa_e$. In addition, a recent work suggests that wavevector-dependent scattering processes, such as boundary scattering (on crystalline boundaries in that case) could also have an increasing impact on $\kappa_l$ as temperature increases.[49]

Finally, we note that in this study there are a number of approximations we have made that we would like to elaborate on. Real material bandstructures include a wide variety of effective masses, band gaps, degeneracies, non-parabolicity and multiple valence and/or conduction bands, band changes upon spin-orbit coupling consideration, etc., and these features can also be temperature dependent. In this study however, we only aim to show to first-order the generic potential for NIs to reduce the bipolar effect in narrow bandgap semiconductors, and we assumed a bandgap value comparable to $Bi_2Te_3$ to relate to common TE materials. Examining further bandstructure details would require a tremendous amount of simulations and is out of the scope of this work. However, to investigate the resilience of our results to changes in the NI barrier height, in Fig. 10 we show (a) the percentage decrease in $\kappa_{bi}$, (b) the percentage decrease in $\kappa_{el,tot}$, and (c) the percentage increase in PF for a bandstructure with masses $m_c = m_0$, $m_v = m_0$, and bandgap of $E_g = 0.2$ eV, with NIs of barrier height $V_B = 0.2$ eV (green-cross lines – same as green-cross lines in Fig. 5), and a bandstructure with the same parameters except NIs barrier heights of $V_B = 1$ eV (green-dashed lines). It can be seen that this causes some additional decrease in $\kappa_{el,tot}$, and increase in *PF*. However, despite the five-time increase in NI barrier height, these additional benefits are only modest. This suggests that any NI material that causes a barrier height of at least $V_B = 0.2$ eV and above, is already enough to limit the hole flow, and that variations in this variable will not have a significant impact on our results.

In order to also observe the influence of a larger bandgap, we also consider an additional bandstructure with masses the same as the green-cross line ($m_c = m_0$, $m_v = m_0$) and NIs of height $V_B = 0.2$ eV, but with an increased bandgap of $E_g = 0.3$ eV (purple cross lines in Fig. 10). This bandgap value is similar to that of PbTe – another important thermoelectric material, for example. It can be seen by comparing the green-cross lines ($E_g = 0.2$ eV) and the purple-cross lines ($E_g = 0.3$ eV) that the NIs have lesser impact at larger bandgaps, although it should be noted that the bipolar effect only becomes significant at higher temperatures in the case of larger bandgap materials. Thus, the same qualitative features and performance improvements will be expected, but at higher temperatures.

Finally, we also note that we have assumed perfect band alignment between the matrix material and the NIs in the conduction band as a best case scenario. In reality, there will always be some scattering added for the conduction band as well when NIs are introduced into the matrix material and this could limit thermoelectric performance benefits at high NI densities, by limiting the overall *G*.

## IV. Conclusions

In conclusion, in this work, using Non-Equilibrium Green's function simulations for electronic transport, we have demonstrated the ability of nanoinclusions, an important component in the design of advanced thermoelectric materials, to reduce the bipolar effect in narrow bandgap materials. By placing nanoinclusions separated on the order of the mean free path of the minority carrier, or even more densely, we showed and quantified how this leads to reductions in the electronic thermal conductivity, but also increases in thermoelectric power factor. These benefits are most pronounced in materials in which the mean-free-path of the minority carriers is large to begin with, either due to low effective masses or low scattering rates. Importantly, however, we showed that in the case of materials with light minority carrier mass (and long mean-free-path) in combination with heavy majority masses, the benefits are even higher, especially for the power factor. In the latter case in particular, the benefits in the power factor are much larger compared to when both majority and minority bands are light. Benefits from nanoinclusions on the power factor also seem more and more significant as temperature increases while benefits on the total electronic thermal conductivity also increase with temperature, but begin to saturate at some point (although the effect of NIs in reducing the bipolar thermal conductivity is lessened with increasing temperature). Finally, we showed that nanoinclusion barriers of a few hundred meV are enough to provide sufficient benefits, whereas higher barriers do not obscure minority carrier transport to a significantly greater degree.

Acknowledgements: This work has received funding from the European Research Council (ERC) under the European Union's Horizon 2020 Research and Innovation Programme (Grant Agreement No. 678763). The authors would like to thank Dr Byungki Ryu for useful discussions and Dr Mischa Thesberg for help in the development of the simulator.

Data availability: The raw data required to reproduce these findings cannot be shared at this time due to technical or time limitations. The processed data required to reproduce these findings cannot be shared at this time due to technical or time limitations.

Figure 1:

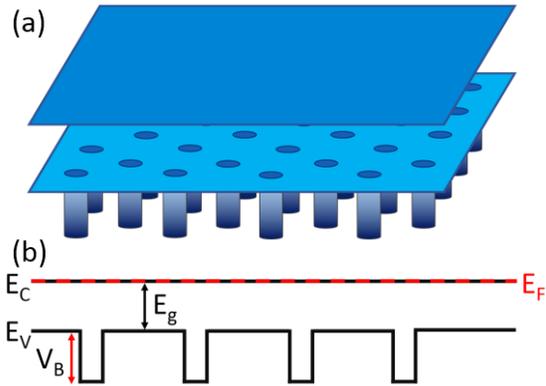

## Figure 1 caption:

(a) A schematic of the potential energy in the 2D channel considered. Nanoinclusions are modelled as cylindrical potential barriers for the minority carriers in the valence band, but do not affect the majority carriers in the conduction band. (b) A 1D schematic identifying the key parameters of the system: the conduction band minimum, $E_C$; the valence band maximum, $E_V$; the bandgap, $E_g$; the nanoinclusion barrier height, $V_B$; and the Fermi level, $E_F$, which here is aligned with the $E_C$.

Figure 2:

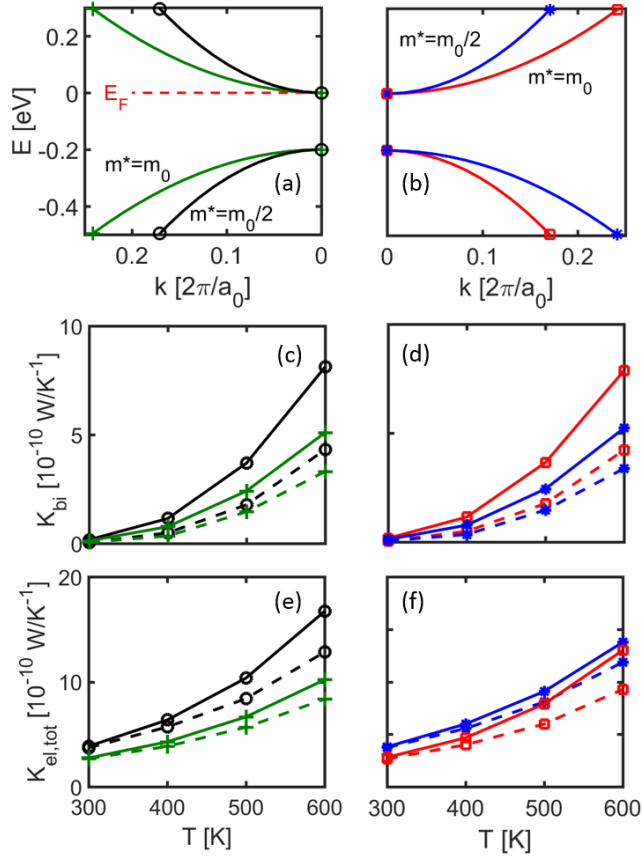

Figure 2 caption:

(a) The symmetrical bandstructures: $m_c = m_0$, $m_v = m_0$ (green-cross lines); and $m_c = 0.5m_0$, $m_v = 0.5m_0$ (black-circle lines). (b) The asymmetrical bandstructures: $m_c = m_0$, $m_v = 0.5m_0$ (red-square lines); and $m_c = 0.5m_0$, $m_v = m_0$ (blue-star lines). (c) The bipolar thermal conductivity versus temperature for the symmetrical bandstructures. (d) The bipolar thermal conductivity versus temperature for the asymmetrical bandstructures. (e) The total electronic thermal conductivity versus temperature for the symmetrical bandstructures. (f) The total electronic thermal conductivity versus temperature for the asymmetrical bandstructures. Results are shown for structures with (dashed lines) and without (solid lines) nanoinclusions.

Figure 3:

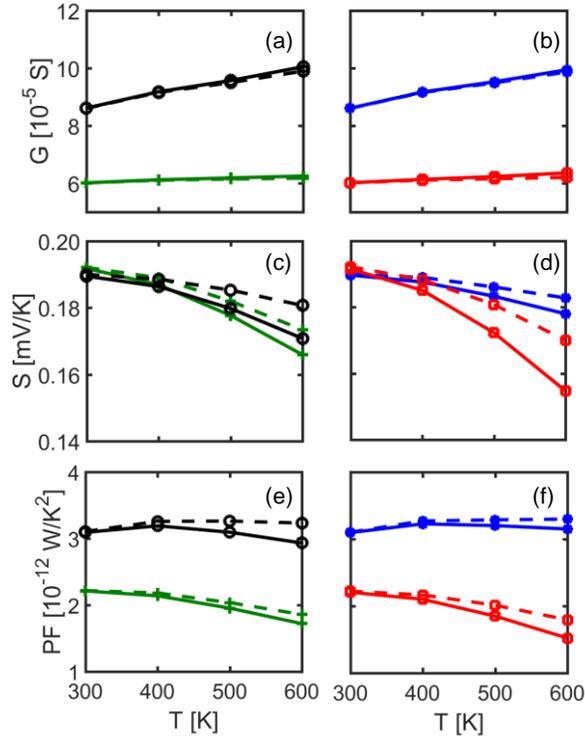

## Figure 3 caption:

The conductance, $G$, Seebeck coefficient, $S$, and power factor, $PF = GS^2$, versus temperature, $T$, for the four bandstructures shown in Fig. 2(a-b): $m_c = m_0$, $m_v = m_0$ (green-cross lines); $m_c = m_0$, $m_v = 0.5m_0$ (red-square lines); $m_c = 0.5m_0$, $m_v = m_0$ (blue-star lines); $m_c = 0.5m_0$, $m_v = 0.5m_0$ (black-circle lines). Results are shown for structures with (dashed lines) and without (solid lines) nanoinclusions.

Figure 4:

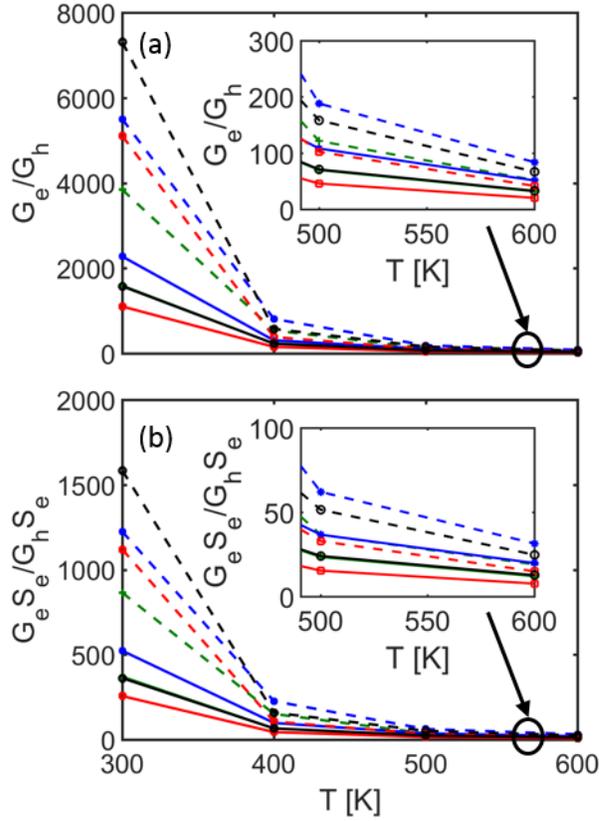

Figure 4 caption:

(a) The ratio of the conduction band conductance, $G_e$, and the valence band conductance, $G_h$ versus temperature. (b) the ratio $G_e S_e / G_h S_h$ versus temperature. Results are for the four bandstructures shown in Fig. 2(a-b): $m_c = m_0$, $m_v = m_0$ (green-cross lines); $m_c = m_0$, $m_v = 0.5 m_0$ (red-square lines); $m_c = 0.5 m_0$, $m_v = m_0$ (blue-star lines); $m_c = 0.5 m_0$, $m_v = 0.5 m_0$ (black-circle lines), and for structures with (dashed lines) and without (solid lines) nanoinclusions.

Figure 5:

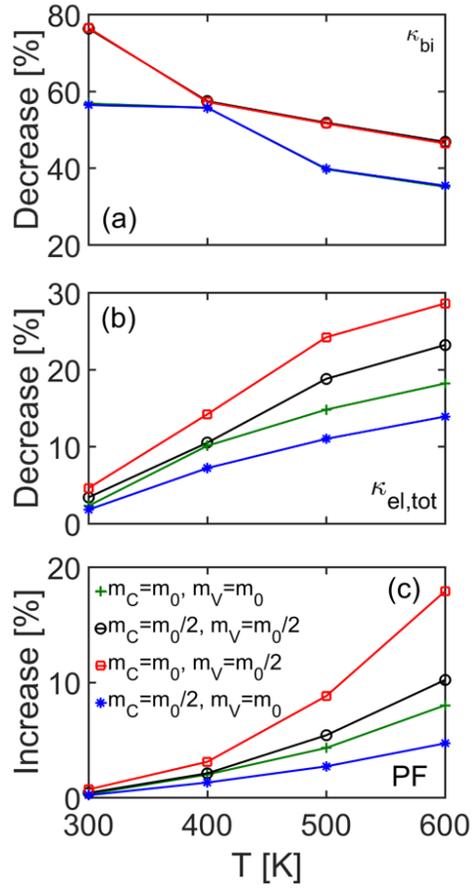

Figure 5 caption:

(a) The percentage decrease in the bipolar thermal conductivity $\kappa_{bi}$ due to the introduction of nanoinclusions on the valence band versus temperature for the four bandstructures shown in Fig. 2(a-b): $m_c = m_0$, $m_v = m_0$ (green-cross lines); $m_c = m_0$, $m_v = 0.5m_0$ (red-square lines); $m_c = 0.5m_0$, $m_v = m_0$ (blue-star lines); $m_c = 0.5m_0$, $m_v = 0.5m_0$ (black-circle lines), (b) The percentage decrease in the total electronic thermal conductivity $\kappa_{el,tot}$ versus temperature. (c) The percentage increase in power factor $PF = GS^2$ versus temperature.

Figure 6:

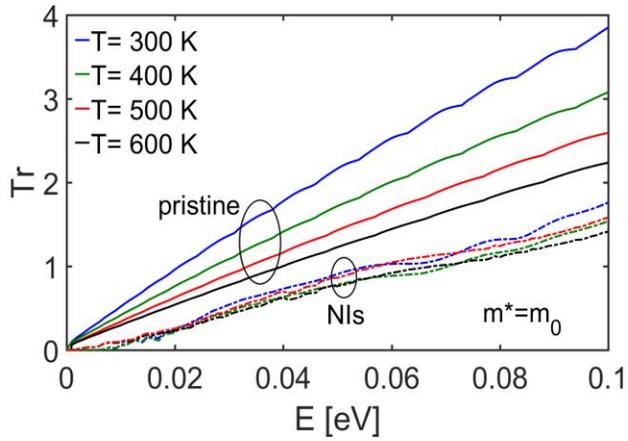

Figure 6 caption:

The transmission versus energy of the $m_v = m_0$ valence band shown in Fig. 2(a) at four temperatures: 300 K (blue lines), 400 K (green lines), 500 K (red lines), 600 K (black lines); and for with (dashed-dotted lines) and without (solid lines) nanoinclusions. Note that the energy values are relative to the valence band edge and increase in value moving into the band.

Figure 7:

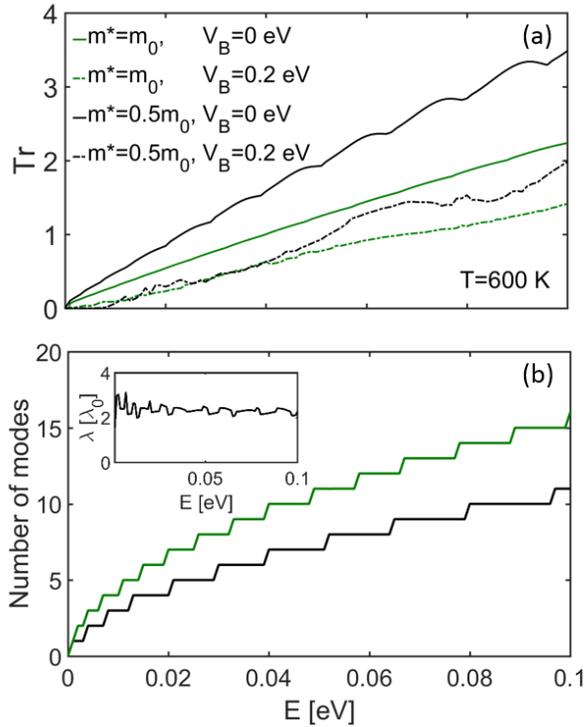

Figure 7 caption:

(a) The transmission versus energy of the two valence bands shown in Fig. 2(a): $m_v = m_0$ (green lines); and $m_v = 0.5m_0$ (black lines). Results are shown for structures with (dashed-dotted lines) and without (solid lines) nanoinclusions. Note that the energy values are relative to the valence band edge and increase in value moving into the band. (b) The number of modes versus energy of those two valence bands. Inset: the mean-free-path of the light band as a proportion of the heavy band mean-free-path versus energy.

Figure 8:

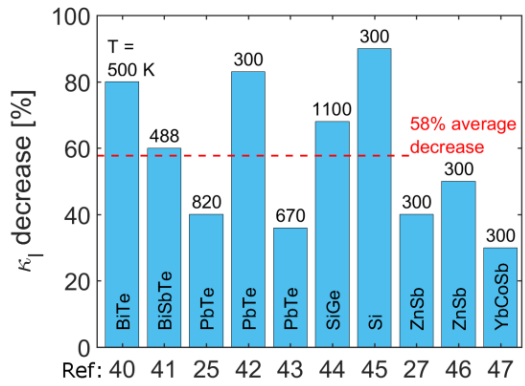

Figure 8 caption:

The percentage decrease in lattice thermal conductivity due to the introduction of nanoinclusions in a selection of materials at various temperatures taken from the literature. Reference numbers are given on the x axis, and the bulk material and temperature at which the reduction was recorded are given within the graph.

Figure 9:

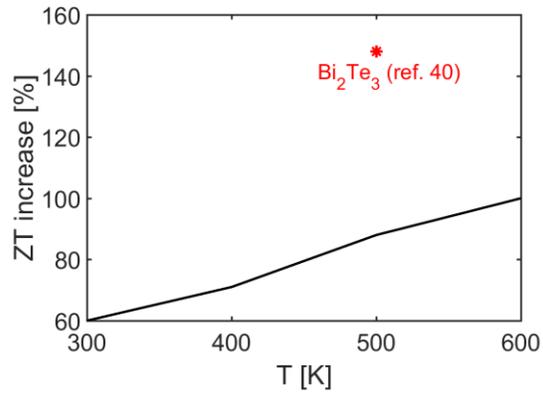

Figure 9 caption:

The percentage increase in *ZT* due to the introduction of NIs versus temperature assuming a constant 58% reduction in $\kappa_l$ (black line) and in addition for the $Bi_2Te_3$ taking the more representative values at 500 K from Ref. 40 (red star).

Figure 10:

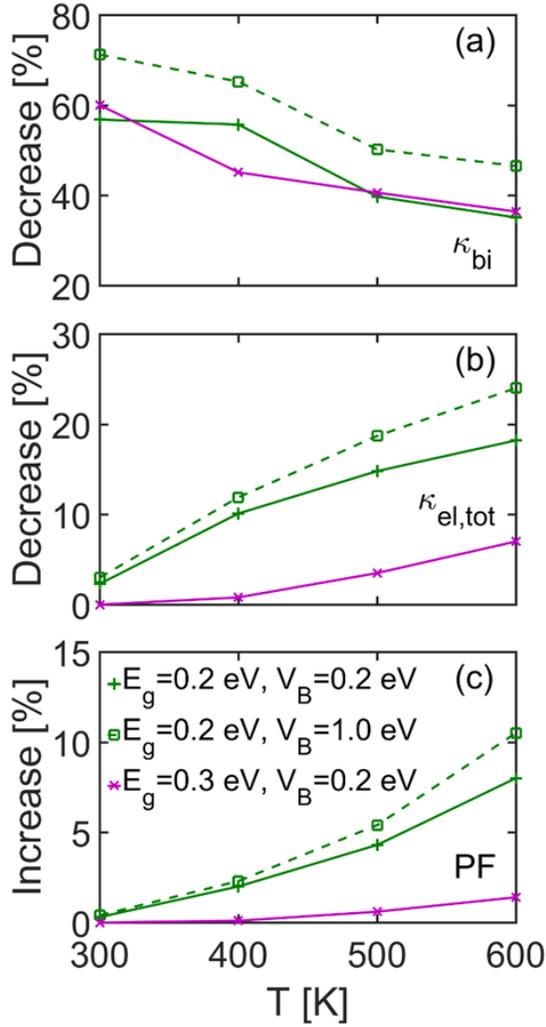

Figure 10 caption:

(a) The percentage decrease in the bipolar thermal conductivity $\kappa_{bi}$ due to the introduction of nanoinclusions on the valence band versus temperature for three bandstructures: $m_c = m_0$, $m_v = m_0$ (green-cross lines) with $E_g = 0.2$ eV and nanoinclusions of height $V_B = 0.2$ eV (the same lines as shown in Fig. 5), a bandstructure with masses the same as the green-cross line ($m_c = m_0$, $m_v = m_0$), but with an increased bandgap of $E_g = 0.3$ eV (purple-cross lines), and a bandstructure with masses the same as the green-cross line ($m_c = m_0$, $m_v = m_0$), but nanoinclusions of height $V_B = 1$ eV (green-dashed lines). (b) The percentage decrease in the total electronic thermal conductivity $\kappa_{el,tot}$ versus temperature. (c) The percentage increase in power factor $PF = GS^2$ versus temperature.

29